\documentclass[tbtags]{article}
\usepackage{amsmath,amssymb,mathrsfs,epsfig,subfigure}
\newcommand{\lagr}{\mathscr{L}}
\newcommand{\partialvob}{\partial\hspace{-6pt}\raisebox{9pt}[0pt]{$\scriptscriptstyle\leftrightarrow$}}
\newcommand{\msbar}{\hspace{0.4mm}\overline{\text{\hspace{-0.4mm}MS\hspace{-0.6mm}}}\hspace{0.6mm}}
\newcommand{\GeV}{\text{GeV}}
\newcommand{\fI}{f^{\Lambda}_{i}}
\newcommand{\fB}{f^{\Lambda}_{\scriptscriptstyle B}}
\newcommand{\fBW}{f^{\Lambda}_{\scriptscriptstyle BW}}
\newcommand{\fW}{f^{\Lambda}_{\scriptscriptstyle W}}
\newcommand{\fDW}{f^{\Lambda}_{\scriptscriptstyle DW}}
\newcommand{\fWW}{f^{\Lambda}_{\scriptscriptstyle WW}}
\newcommand{\fWWW}{f^{\Lambda}_{\scriptscriptstyle WWW}}
\newcommand{\fBB}{f^{\Lambda}_{BB}}
\begin{document}
\vspace*{2cm}
\begin{center}
{\Large {\bf Contribution of dimension-six bosonic operators to $H\rightarrow\gamma\gamma$ at one-loop level}}\\[0.6cm]
Ji\v{r}\'{\i}
Ho\v{r}ej\v{s}\'\i\footnote{Jiri.Horejsi@mff.cuni.cz},
Karol Kampf\footnote{karol.kampf@mff.cuni.cz}\\[0.3cm]
Institute of Particle and Nuclear Physics,\\ Faculty of Mathematics and Physics, Charles University\\
V Hole\v{s}ovi\v{c}k\'ach 2, CZ-18000 Prague, Czech Republic
\end{center}\vspace{1.5cm}
\begin{abstract}
The decay process $H\rightarrow\gamma\gamma$ is examined in a
model-independent way within the effective Lagrangian approach.
Contribution of a set of irreducible one-loop diagrams involving
$SU(2)\times U(1)$ invariant dimension-six bosonic operators is
evaluated explicitly. The calculation is intended to fill a gap
that exists in the current literature on the subject.
\\[0.5cm]
{\it keywords\/}: electroweak interactions; effective Lagrangians;
Higgs decay into two photons.
\\[0.3cm]
\end{abstract}
\newpage
\section{Introduction}
The standard model (SM) of fundamental interactions has so far
proved to be a very successful theory, at least from the point of
view of present-day phenomenology. The most prominent unsolved
issue is the mechanism of the electroweak symmetry breaking (EWSB)
and a possible existence of the related Higgs scalar sector.
Current precision tests of the SM seem to point towards a
relatively light Higgs boson, i.e. such that $m_H<200\,\text{GeV}$
or so (see e.g.~\cite{barate03}). Needless to say, such an
``evidence" is only indirect, as it is based on a successful fit
of calculated effects of the relevant closed-loop Feynman graphs
to the precision electroweak data. Note also that there is another
message from the EWSB sector, which is more general and rather
spectacular. For the famous parameter
$\rho=m_W^2/m_Z^2\cos^2\theta_W$ (with $m_W$ and $m_Z$ denoting
the vector boson masses and $\theta_W$ being the weak mixing
angle) the experimental data show that, with great accuracy,
$\rho=1$. Such a value is naturally obtained from the Higgs sector
with doublet structure, or, more generally, from an EWSB mechanism
respecting the custodial $SU(2)$ symmetry~\cite{sikivie80}.

On the other hand, a general opinion prevailing nowadays is that
SM cannot be, for various reasons, an ultimate model of particle
physics -- in other words, it should be understood as an effective
theory valid below the energy scale of the order
$O(1\,\text{TeV})$ and somewhere above that scale, some kind of
new physics is to be expected (for a review, see e.g.~\cite{bsm}).
A general framework for describing the effects of the physics
beyond SM is an appropriate effective Lagrangian. This
incorporates, besides the standard renormalizable interaction
terms with dimension four and three, also non-renormalizable
higher-dimensional terms involving negative powers of a $\Lambda$,
the relevant scale of new physics. Thus, the generic form of such
an effective electroweak Lagrangian reads
\begin{equation}\label{Leff}
\lagr_{\text{\it eff}} = \lagr_{SM} +
\sum_{n\geq5}\sum_{i=1}^{N_n} \frac{\alpha_i^{(n)}}
{\Lambda^{n-4}} \mathcal{O}_i^{(n)}.
\end{equation}
When constructing such an extension of the SM, gauge invariance
under $SU(2)\times U(1)$ is required and, in addition to that, one
can make a specific assumption concerning the $\Lambda$. We shall
assume that $\Lambda\gg v$, where $v\doteq 246\,\text{GeV}$ is the
usual electroweak scale. This corresponds to the so-called
``decoupling scenario", in which the $SU(2)\times U(1)$ symmetry
is realized linearly (concerning the terminology, see
e.g.~\cite{wudka94}, \cite{ns} and references therein). In
particular, this means that the unphysical Goldstone bosons enter
via the standard complex $SU(2)$ doublet, along with the
elementary Higgs boson field and, generally, the low-energy
spectrum is supposed to be identical with that of SM. Explicit
representation of the non-renormalizable higher-order terms
invariant under $SU(2)\times U(1)$ gauge symmetry can be found in
the literature, see in particular~\cite{buchmuller85},
\cite{hagiwara93}. Contributions of the higher-dimensional terms
to the amplitudes of low-energy processes are then in fact
suppressed by powers of the ratio $v/\Lambda$.

In future experimental searches for the Higgs scalars, the process
$H\rightarrow\gamma\gamma$ might play an important role, since a
light scalar boson (i.e. that with $m_H<2m_W$) could be detected
through this decay mode at LHC (cf.~\cite{cranmer04} and
references therein). Of course, there is no tree-level coupling
$H\gamma\gamma$ within SM: the lowest-dimensional interaction term
of this type is necessarily proportional to
$HF_{\mu\nu}F^{\mu\nu}$ (where $F_{\mu\nu}$ denotes the
electromagnetic field strength) if one is to maintain the
electromagnetic gauge invariance. However, such a monomial has
obviously dimension five and is therefore non-renormalizable.
Consequently, SM yields a calculable (i.e. ultraviolet-finite)
result for $H\rightarrow\gamma\gamma$ matrix element at one-loop
level and this was obtained long ago~\cite{ellis75}. It can also
be sensitive to possible effects of the new physics described
schematically by the expansion~(\ref{Leff}) and in current
literature there are several papers dealing with this issue (see
e.g.~\cite{konig91}, \cite{hgg}, \cite{ns}); let us remark that
contributions of higher-dimensional effective interactions to the
inverse process $\gamma\gamma \rightarrow H$ have also been
considered previously~\cite{banin98}. Nevertheless, what seems to
be still missing, is a one-loop calculation involving a complete
set of $SU(2)\times U(1)$ invariant operators that would yield the
leading correction (in powers of $1/\Lambda$) to the SM result.
Some important steps toward this goal have already been made
previously in the papers~\cite{konig91}, \cite{hgg} and~\cite{ns}.
In the present paper we would like to fill a gap that, in our
opinion, occurs in the existing treatments. In particular, our
effort is concentrated on the part of the calculation that is
perhaps most tedious from the technical point of view: we evaluate
one-particle irreducible purely bosonic one-loop graphs involving
just one insertion of a dimension-six effective interaction vertex
from~(\ref{Leff}), with other ingredients descending from SM. An
important consistency check of our computation is the transverse
tensor structure of the matrix element -- this is achieved,
similarly as in the SM case, only when a set of relevant graphs is
summed. Thus, the results presented in this paper are of rather
technical nature, but we believe that they may prove useful in a
future complete calculation based on the $SU(2)\times U(1)$
effective Lagrangian of the type~(\ref{Leff}).

The paper is organized as follows. In the next section we give a
brief summary of the basic SM results for $H\rightarrow
\gamma\gamma$. In Section 3, the set of $SU(2)\times U(1)$ bosonic
operators with dimension six is shown explicitly and in Section 4
we display the main features of the whole calculation. The results
are described in some detail in Section 5 and Section 6 contains
their short summary and some concluding remarks. To make the paper
self-contained, we have added the Appendix containing the
definitions and explicit analytic formulae for the
Passarino-Veltman functions that are necessary for the evaluation
of loop integrals.
\newpage
\section{Standard Model Results}
\label{SMp} To begin with, let us reproduce here -- for reference
purposes -- the form of the electroweak SM Lagrangian. In the
$U$-gauge it reads (for a concise summary see
e.g.~\cite{horejsi02})
\begin{equation}\begin{split}
\lagr^{(GWS)}_{int} &= \sum_f Q_f e \bar{f} \gamma^\mu f A_\mu +
\lagr_{CC} + \lagr_{NC}\\ &+ig (W^0_\mu W^-_\nu \partialvob^\mu
W^{+\nu} + W^-_\mu W^+_\nu \partialvob^\mu W^{0\nu} + W^+_\mu
W^0_\nu \partialvob^\mu W^{-\nu})\\
&-g^2 \bigl[ \tfrac{1}{2}(W^-\cdot W^+)^2 - \tfrac{1}{2}(W^-)^2
(W^+)^2 + (W^0)^2 (W^-\cdot W^+) \\ &\hspace{5.7cm}- (W^-\cdot
W^0)(W^+\cdot W^0) \bigr]\\
&+gm_W W^-_\mu W^{+\mu} H + \frac{1}{2\cos\theta_W} g m_Z Z_\mu
Z^\mu H \\ &+\frac{1}{4} g^2 W^-_\mu W^{+\mu} H^2 +
\frac{1}{8}\frac{g^2}{\cos^2\theta_W} Z_\mu Z^\mu H^2 \\
&-\sum_f \frac{1}{2} g \frac{m_f}{m_W} \bar{f}f H - \frac{1}{4} g
\frac{m_H^2}{m_W} H^3 - \frac{1}{32} g^2 \frac{m_H^2}{m_W^2} H^4,
\end{split}\end{equation}
where we have denoted
\begin{equation} W^0_\mu = \cos\theta_W Z_\mu + \sin\theta_W
A_\mu.\label{Lin1}
\end{equation}
The $\lagr_{CC}$ and $\lagr_{NC}$ are the usual symbols for
charged and neutral current interactions of fermions with vector
fields; an explicit form of these terms will not be needed for our
calculations.

\begin{figure}[ht]
\begin{center}
\subfigure[]{\scalebox{0.88}{\epsfig{file=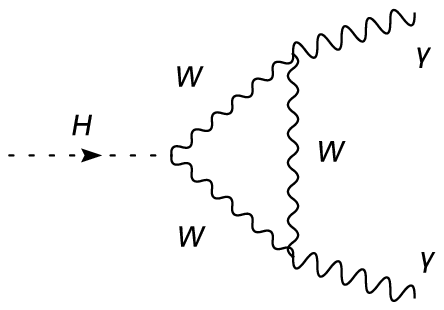}}}
\subfigure[]{\scalebox{0.88}{\epsfig{file=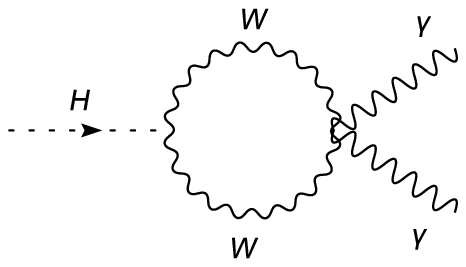}}}\subfigure[]{\scalebox{0.88}{\epsfig{file=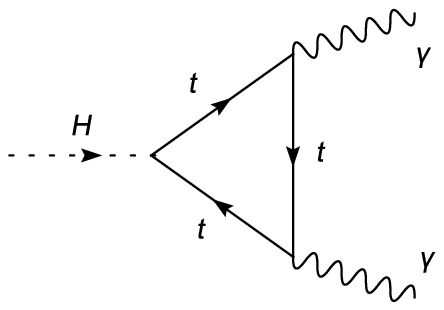}}}
\end{center}
\caption{SM one-loop diagrams for $H\rightarrow\gamma\gamma$.
Crossing of the external photon lines is implicit in contributions
of the graphs (a) and (c).} \label{hgg1}
\end{figure}

As we have already stressed in the Introduction, for obvious
dimensional reasons there is no direct tree-level $H\gamma\gamma$
coupling within the SM Lagrangian. The one-loop SM graphs for the
decay process $H\rightarrow\gamma\gamma$ are shown in
Fig.\ref{hgg1}.
Note that at this level, the fermionic part is dominated by the
contribution of the heaviest fermion species. Therefore we
consider here the top quark loop only.

A straightforward calculation yields the following result for the
decay matrix element:
\begin{equation}\label{MSM}
{\cal M} = (F_\text{VB}+F_\text{top}) \Bigl(g^{\alpha\beta} -
\frac{2 k^\alpha p^\beta}{m_H^2} \Bigr) \varepsilon^*_\alpha(p)
\varepsilon^*_\beta(k),
\end{equation}
where the ``formfactors" $F_\text{VB}$ and $F_\text{top}$
represent the contributions of the vector boson and top quark loop
respectively. These are expressed through
\begin{align}
F_\text{VB} &= \frac{3 i m_W}{8\pi^2}\bigl[1 + \frac{m_H^2}{6
m_W^2} + (2 m_W^2 - m_H^2)
C_0(m_H^2,0,0,m_W^2,m_W^2,m_W^2)\bigr]\,, \label{Fbos}\\
F_\text{top} &= \frac{-i}{6\pi^2} \frac{m_t^2}{m_W} \bigl[ 2 + (4
m_t^2-m_H^2) C_0(m_H^2,0,0,m_t^2,m_t^2,m_t^2)\bigr]\,,\label{Ffer}
\end{align}
with $C_0$ denoting the relevant Passarino-Veltman function (see
e.g.~\cite{bardin99}); the definitions and explicit analytic
formulae for a set of these functions are summarized in the
Appendix.

Notice also that in writing the result~(\ref{MSM}), we have
pinpointed the tensorial factor
\begin{equation}\label{Tmn}
T^{\mu\nu} = g^{\mu\nu} - 2 m_H^{-2}k^\mu p^\nu,
\end{equation}
which is transverse with respect to the external photon momenta,
i.e.
\begin{equation}
p_\mu T^{\mu\nu} = 0,\qquad k_\nu T^{\mu\nu}=0 \label{T0}
\end{equation}
as anticipated on the basis of the electromagnetic gauge
invariance. The appearance of the transverse tensor
structure~(\ref{Tmn}) will serve as an important consistency check
also in our subsequent calculations with the effective
Lagrangian~(\ref{Leff}).

Finally, let us add that from the matrix element~(\ref{MSM}) one
obtains the decay width
\begin{equation}
\label{GSM} \Gamma_\text{SM}(H\rightarrow \gamma\gamma) =
\frac{8\pi\alpha^2 G_F}{\sqrt{2}}\frac{m_W^2}{m_H} | F_\text{top}
+ F_\text{VB} |^2 \, ,
\end{equation}
(cf.~\cite{bardin99}).

\section{Dimension-Six Effective Operators}\label{Section3}
Let us now proceed to examine in more detail the effective
interaction terms with dimension greater than four, described
schematically by the second part of the Lagrangian~(\ref{Leff}).
As noted in the Introduction, we shall restrict ourselves to the
dimension-six bosonic operators. Thus, the effective Lagrangian we
are going to work with has a general form
\begin{equation}\label{Le6}
\lagr_{\text{\it eff}} = \lagr_{SM} + \sum_i
\frac{f_i^{(6)}}{\Lambda^2} \mathcal{O}_i^{(6)}.
\end{equation}
For the moment, let us leave aside any discussion of possible
``realistic" values of the mass scale $\Lambda$ and coupling
constants $f_i^{(6)}$ that parameterize the effects of a new
physics beyond SM. Here we only give a full list of the operators
$O_i^{(6)}$ that will be utilized in our calculation. An explicit
construction of such a set was studied systematically e.g.
in~\cite{hagiwara93}. Assuming that the bosonic sector of our
effective low-energy theory contains only the particles $W^\pm$,
$Z$, $\gamma$ and $H$, and imposing the requirement of
$SU(2)\times U(1)$ gauge symmetry together with separate
invariance under charge conjugation and parity, one arrives at
eleven independent dimension-6 monomials in the relevant field
variables, namely
\begin{eqnarray}
&& {\cal O}_{WWW}=\text{Tr}[\hat{W}_{\mu
\nu}\hat{W}^{\nu\rho}\hat{W}_{\rho}^{\mu}]
, \nonumber \\
&& {\cal O}_{WW} = \Phi^{\dagger} \hat{W}_{\mu \nu}
\hat{W}^{\mu \nu} \Phi , \nonumber \\
&&{\cal O}_{BW} =  \Phi^{\dagger} \hat{B}_{\mu \nu}
\hat{W}^{\mu \nu} \Phi, \nonumber \\
&& {\cal O}_{DW} = \text{Tr}\bigl([ D_\mu ,\hat{W}_{\nu \rho}]
[ D^\mu ,\hat{W}^{\nu\rho} ]\bigr), \nonumber \\
&&{\cal O}_{DB}=-\frac{{g^\prime}^2}{2} \left(\partial_\mu
B_{\nu\rho}\right) \left(\partial^\mu
B^{\nu\rho}\right),\nonumber \\
&&{\cal O}_{BB} = \Phi^{\dagger} \hat{B}_{\mu \nu} \hat{B}^{\mu
\nu} \Phi ,  \nonumber \\
&&{\cal O}_W  = (D_{\mu} \Phi)^{\dagger}
\hat{W}^{\mu \nu}  (D_{\nu} \Phi), \nonumber \\
&&{\cal O}_B  =  (D_{\mu} \Phi)^{\dagger} \hat{B}^{\mu \nu}
(D_{\nu} \Phi), \nonumber \\
&&{\cal O}_{\Phi,1} = \left ( D_\mu \Phi \right)^\dagger
\Phi^\dagger \Phi
\left ( D^\mu \Phi \right ), \nonumber \\
&&{\cal O}_{\Phi,2} =\frac{1}{2}
\partial^\mu\left ( \Phi^\dagger \Phi \right)
\partial_\mu\left ( \Phi^\dagger \Phi \right), \nonumber \\
&&{\cal O}_{\Phi,3} =\frac{1}{3} \left(\Phi^\dagger \Phi
\right)^3. \label{c6}
\end{eqnarray}
Some explanatory remarks concerning the notation employed
in~(\ref{c6}) are in order here. For convenience, we have
introduced the symbols $\hat{B}$ and $\hat{W}$ for vector field
variables including the gauge coupling constants:
\begin{equation}
\hat{B}_{\mu\nu} = \frac{1}{2}ig' B_{\mu \nu},\hspace{1cm}
\hat{W}_{\mu\nu} = i g \frac{\sigma^a}{2} W^a_{\mu \nu},
\end{equation}
with $\sigma^a$ being the standard Pauli matrices and
\begin{align}
B_\mu &= -\sin\theta_W Z_\mu + \cos\theta_W A_\mu\\
W^\mp_\mu &= \frac{1}{\sqrt{2}}(W^1_\mu \pm i W^2_\mu).
\end{align}
The covariant derivative acting on the isospin doublet $\Phi$ has
the form
\begin{equation}
D_\mu = \partial_\mu + \frac{i}{2}g' B_\mu + i
g\frac{\sigma^a}{2}W^a_\mu.
\end{equation}
 The Higgs boson field is introduced by means of the $\Phi$ in the usual way; in the $U$-gauge this becomes
\begin{equation}
\Phi = \frac{1}{\sqrt{2}}\begin{pmatrix}0\\ v + H \end{pmatrix}.
\label{PhiU}
\end{equation}

\section{Contribution to $H\rightarrow\gamma\gamma$ from $\lagr^{(6)}_\text{boson}$}

With the basic SM results for the $H\rightarrow\gamma\gamma$ decay
amplitude at hand (see (\ref{MSM})-(\ref{Ffer})), a natural next
goal is to calculate the corresponding leading correction due to
the $O(1/\Lambda^2)$ terms in the effective Lagrangian
(\ref{Le6}). To this end, we shall consider tree-level and
one-loop diagrams, in which a vertex originating in the
$\mathcal{L}_\text{boson}^{(6)}$ appears exactly once. As we have
stated in the Introduction, we are primarily interested in a
complete evaluation of the one-particle irreducible one-loop
bosonic graphs. Note that for this type of contributions, the
transversality expressed by eq.~(\ref{T0}) represents a
non-trivial criterion of correctness of the whole calculation, as
it is achieved only in the sum of all relevant diagrams (in
contrast to that, a contribution of any individual reducible
diagram is transverse by itself).

Two remarks are in order here. First, the complete calculation
would have to include an appropriate renormalization procedure: UV
divergent parts of one-loop graphs, evaluated by means of
dimensional regularization, should be absorbed into a redefinition
of the tree-level coefficients of the effective Lagrangian -- in
other words, they are to be cancelled by appropriate counterterms,
whose structure is determined by the considered
Lagrangian~(\ref{Le6}). In the present paper, we do not implement
such a full-fledged renormalization procedure, since we
temporarily ignore some contributions that have been studied
previously by other authors (for example, contributions due to
dimension-six fermionic operators and, in general, contributions
of various reducible graphs). We prefer to give here just a
separate treatment of the above-mentioned irreducible graphs (as
far as we know, these have not been calculated before) and a more
complete discussion will be presented elsewhere. For practical
purposes, when displaying the results obtained for the diagrams in
question (see the next section), we remove the UV divergent parts
according to the $\msbar$ scheme.

The second remark concerns the structure of the
operators~(\ref{c6}). Taking into account the $U$-gauge
representation of the Higgs doublet~(\ref{PhiU}), one can notice
that e.g. the $O_{WW}$ and $O_{BB}$ contain, among other things,
also parts proportional to the vector boson kinetic terms (of
course, these are obtained simply by replacing the $\Phi$ with its
vacuum expectation value). In our calculation, contribution of
such bilinear terms will be included in the considered Feynman
graphs in a straightforward manner, i.e. as any other
dimension-six effective coupling. An alternative approach would
consist in incorporating such effects into the renormalization
constants of the SM (cf. e.g.~\cite{hagiwara93}). Note that
bilinear interaction terms arise also from the other
operators~(\ref{c6}).

\begin{figure}[ht]
\begin{center}
\subfigure[]{\epsfig{file=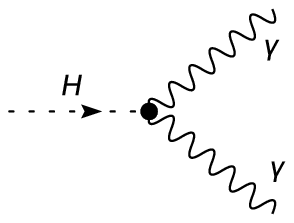}}\qquad\qquad
\subfigure[]{\epsfig{file=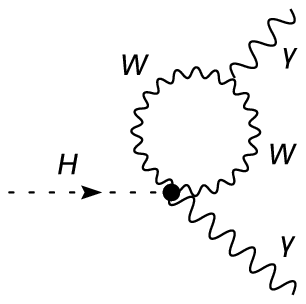}}\\[-12pt]
\subfigure[]{\epsfig{file=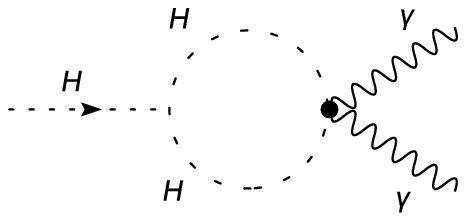}}\qquad\quad
\subfigure[]{\epsfig{file=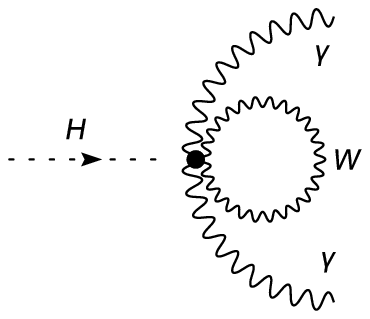}}\vspace{-12pt}
\end{center}
\caption{Irreducible graphs with other than SM-like topologies.
The heavy dot stands for a dimension-six effective vertex}
\label{hggeff}
\end{figure}

Now, let us describe in more detail the set of Feynman graphs that
we are going to calculate. As indicated above, we first take the
graphs in Fig.~\ref{hgg1} and in each of them replace
consecutively a SM vertex with one of those stemming from the
$\lagr_\text{boson}^{(6)}$. Similarly, one can make insertions of
the $O(1/\Lambda^2)$ terms into propagators within the SM graphs.
In this way, one gets 16 diagrams, including the crossing of
external legs. Next, there are several graphs with a new (i.e. not
SM-like) topology. These are depicted in Fig.~\ref{hggeff}
(addition of the crossed diagram to Fig.~\ref{hggeff}b is tacitly
assumed).\vspace{-8pt}

\begin{figure}[ht]
\begin{center}
\epsfig{file=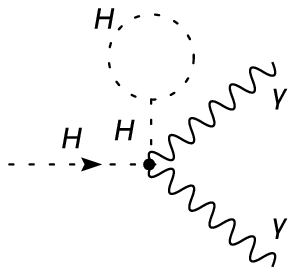}\quad \epsfig{file=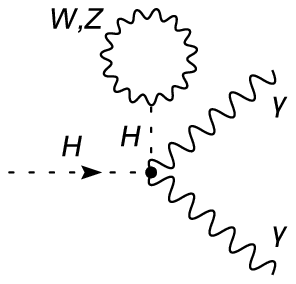}
\end{center}
\caption{$H\rightarrow\gamma\gamma$: tadpole-type vertex
corrections} \label{hggeff2}
\end{figure}

Finally, just for an illustration let us give some examples of
reducible graphs. Two such contributions are shown in
Fig.~\ref{hggeff2}.
As we have stated before, we will not evaluate them explicitly
(though it would be relatively easy). Let us only make the
following remark: Closed loops appearing in Fig.~\ref{hggeff2}
belong to the tadpole-type SM subgraphs, which are known to
produce corrections to the Higgs vacuum expectation value $v$. In
general, such effects can be described in terms of appropriate
running value $v(\mu)$ depending on the relevant renormalization
scale. It turns out (cf.~\cite{arason91}) that corrections due to
running of the $v$ are rather small and thus, in a complete
calculation, one would not make a significant error when using
everywhere the basic value $v(m_W)=246.22\,\text{GeV}$ obtained
from the Fermi constant (measured precisely in muon decay).

\section{Results and Discussion}
Let us now describe the results of our calculation. As we noted
above, the matrix element obtained from the considered Feynman
diagrams exhibits the desired transverse tensor structure:
\begin{equation}\label{MX}
{\cal M}(H\rightarrow \gamma\gamma) = X \Bigl(g^{\alpha\beta} -
\frac{2 k^\alpha p^\beta}{m_H^2} \Bigr) \varepsilon^*_\alpha(p)
\varepsilon^*_\beta(k),
\end{equation}
where $X$ stands for the relevant formfactor. This can be written,
in an obvious notation, as
\begin{equation}
X = e^2 g F_{\text{SM}} + e^2 F^{\text{tree}}_{\text{eff}} + e^2 g
F^{\text{loop}}_{\text{eff}}
\end{equation}
and the corresponding decay rate is then given by
\begin{equation}
\Gamma(H\rightarrow\gamma\gamma) = \frac{8\pi\alpha^2
G_F}{\sqrt{2}}\frac{m_W^2}{m_H} |F_{\text{SM}} + \frac{1}{g}
F^{\text{tree}}_{\text{eff}} + F^{\text{loop}}_{\text{eff}}|^2 \,.
\end{equation}

Since we take into account only bosonic operators in our
evaluation of the leading $O(1/\Lambda^2)$ correction to the SM
result, from now on we will also consider only the contribution of
bosonic loops within SM -- in other words, for the moment we will
ignore fermions altogether. Thus, we will set
\begin{equation}
\label{Floop} F_\text{SM} = F_\text{VB}
\end{equation}
henceforth (cf. (\ref{Fbos})). The contribution of dimension-six
operators~(\ref{c6}) at the tree level has the following simple
form:
\begin{equation}
\label{Ftree} F^\text{tree}_\text{eff} = \frac{i}{2}
(\frac{f_{BB}}{\Lambda^2}-\frac{f_{BW}}{\Lambda^2}+\frac{f_{WW}}{\Lambda^2})
m_H^2 v\,.
\end{equation}
The non-trivial part of our calculation is represented by the
closed-loop contribution. We have five basic topologies (plus two
crossings); in the corresponding diagrams we consider insertions
of effective dimension-six operators -- both in vertices and as
propagator corrections. In total, there are 23 one-loop tensor
integrals to be calculated. Obviously, such a calculation is huge:
a simple counting indicates that one has to expect tensors made of
loop momentum with up to ten indices, schematically
$$
\sim \int d^4 l\; l^{\mu_1}\ldots l^{\mu_{10}} \times
[\text{denominators}]^{-1},
$$
where six $(3\times 2)$ $l^\mu$'s come from propagators, one
$l^\mu$ from SM vertices and three loop momenta descend from an
effective vertex. Thus, the algebraic reduction of these
expressions to the standard scalar integrals is rather long and
tedious. Let us also emphasize that the transversality displayed
in eq.(\ref{MX}) is achieved only when all relevant irreducible
graphs are summed; among those, only Fig.~\ref{hggeff}c  is
transverse by itself.

The main result of this paper is the following formula for the
$F^\text{loop}_\text{eff}$:
\begin{multline}
128 m_W \pi^2 F^\text{loop}_\text{eff} = \\ - 8i\Bigl(\fB - 2 \fBW
+ \fW + 3 g^2 ( -4 \fDW + \fWWW )\Bigr) m_H^2
m_W^2 B_0(0, m_W^2, m_W^2)\\
- 3 (\fBB  - \fBW  + \fWW ) m_H^4 B_0(m_H^2, m_H^2, m_H^2)\\
+2\Bigl\{m_H^2 \Bigl((\fB - 2 \fBW  + \fW ) m_H^2 - 2 (\fB - 2
\fBW + \fW  \\+ 12 g^2 (-4 \fDW  + \fWWW )) m_W^2\Bigr) B_0(m_H^2,
m_W^2,
m_W^2)\\
+ 4 m_W^2 \Bigl[\Bigl(\fB - 2 (\fBW  + 7 g^2 \fDW ) + 3 g^2 \fWWW
\Bigr) m_H^2  - 24 ( g^2 \fDW  + \fWW  ) m_W^2\\ + 2 \Bigl((\fW -
2 \fWW ) m_H^4 - (\fB - 2 \fBW  + 2 (\fW  - 7 \fWW ) + 3 g^2 (-8
\fDW  + \fWWW )) m_H^2 m_W^2 \\- 36 (g^2 \fDW + \fWW ) m_W^4\Bigr)
C_0(m_H^2, 0, 0; m_W^2)\\ + 4 (g^2 \fDW  + \fWW ) m_W^2
\Bigl(-(m_H^2 + 6 m_W^2) C_0(m_H^2, m_H^2, 0; m_W^2)\\ - 2 m_W^2
(-m_H^2 + 6 m_W^2) (D_0(m_H^2; m_W^2) + 2 D_0(m_H^2, m_H^2;
m_W^2))\Bigr)\Bigr]\Bigr\}. \label{fted}
\end{multline}
The $B_0$, $C_0$ and $D_0$ denote the relevant Passarino-Veltman
functions (see Appendix) and we have used a shorthand notation
$\fI = f_i/\Lambda^2$. Note that UV divergences reside only in the
$B_0$ and can be removed by means of the $\msbar$ prescription
which amounts to the replacement
\begin{equation}
f_i = f_i^r + \xi_i \frac{1}{\tilde \varepsilon}\,
\end{equation}
(for the definition of the $1/\tilde\varepsilon$ see Appendix).
Then it is not difficult to realize that the following relation
holds
\begin{multline}
\xi_{BB}-\xi_{BW}+\xi_{WW} = \frac{g^2}{32\pi^2}\bigl(72
f^r_{DW}g^2 -18 f^r_{WWW}g^2 -3 f^r_B + 6 f^r_{BW} -3 f^r_W\\ +
\frac{m_H^2}{4 m_W^2} ( 2 f^r_B - 3 f^r_{BB} -3 f^r_{BW} + 2 f^r_W
-3 f^r_{WW} ) \bigr).
\end{multline}
(at least at the level of the considered irreducible graphs).

Clearly, the result~(\ref{fted}) contains a lot of essentially
unknown free parameters, so it is rather difficult to make any
specific physical prediction for a correction to the SM decay
rate. Anyway, for illustration of our general formula let us
display at least some numerical examples. First, we discuss
briefly the behaviour of the $H\rightarrow\gamma\gamma$ decay
width as a function of the Higgs boson mass.
\begin{figure}[th]
\begin{center}
\epsfig{file=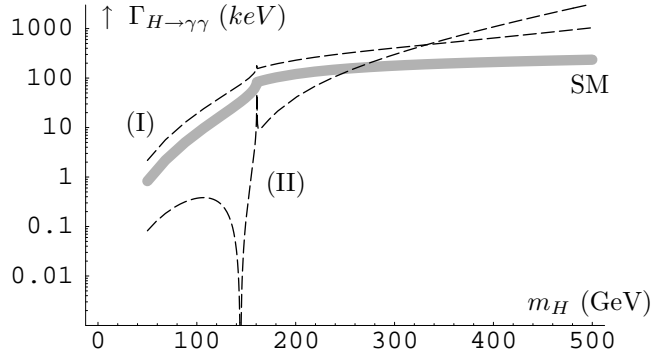} \caption{Bosonic contributions to
$\Gamma(H\rightarrow \gamma\gamma)$ for $\Lambda = 1\,\text{TeV}$
and (I): all $f_i = 1$ (II): $f_{WW}=1.8$, $f_{BB}=-6.5$,
$f_{BW}=-3$, $f_{DW}=5.4$, $f_{WWW}=5.5$, $f_B=-4$ and $f_W=8$.}
\label{gr1}
\end{center}
\end{figure}
In Fig.~\ref{gr1} we have shown two examples of such a dependence,
corresponding to different choices of the $f_i$'s and with
$\Lambda=1\,\text{TeV}$. In the first option, we set all $f_i$'s
equal to 1. The second (more or less random) choice is designed to
demonstrate a possible reduction of the $\Gamma$ in contrast to
the otherwise typical enhancement of the SM result.\footnote{In
fact, the second choice has been set so as to reduce considerably
the SM result (for $m_H\doteq 120\,\text{GeV}$) by means of the
tree-level contribution of dimension-six operators.}

One could also wonder how much is this process dependent upon the
individual constants $f_i$. We could expect that the leading
contribution comes from the tree level (cf. (\ref{Ftree})). It is
of course true, but we can imagine a situation where the
tree-level contribution is suppressed and the one-loop graphs
dominate.
\begin{figure}[th]
\begin{center}
\epsfig{file=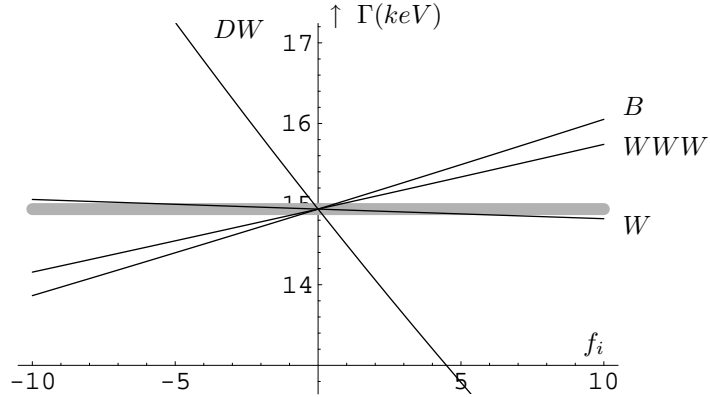} \caption{Bosonic contributions to
$\Gamma(H\rightarrow \gamma\gamma)$: dependence on a given
coefficient $f_i$ with all other being set to zero.
$m_H=120\,\GeV$ and $\Lambda = 1\,\text{TeV}$.} \label{gr2}
\end{center}
\end{figure}
From Fig.~\ref{gr2} we can see that apart from $f_{BB}$, $f_{BW}$
and $f_{WW}$ (which are not considered in this plot because of
their substantial contribution at the tree level), the $f_{DW}$
can give a rather large contribution.

In the previous discussion we have always set $\Lambda=1\,
\text{TeV}$, which is a generic estimate of the new physics scale.
\begin{figure}[th]
\begin{center}
\epsfig{file=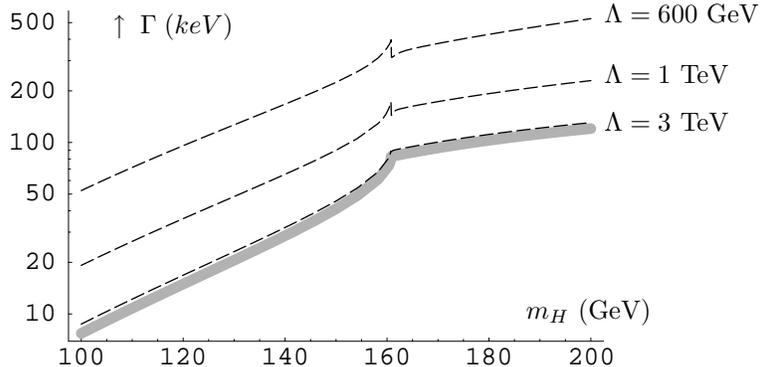} \caption{Bosonic contributions to
$\Gamma(H\rightarrow \gamma\gamma)$: the dependence on $\Lambda$.}
\label{gr3}
\end{center}
\end{figure}
In Fig.~\ref{gr3} we have depicted the dependence of the
calculated width $\Gamma$ on the $\Lambda$. As we would expect,
effects of the new physics are, roughly speaking, the more
important the smaller the scale $\Lambda$ is.
\section{Conclusion}
Using the effective Lagrangian approach, we have examined the
potentially interesting rare decay of the SM-like Higgs boson into
two photons. We have considered the ``decoupling scenario", in
which the onset of a new physics beyond SM is supposed to be
characterized by a mass scale $\Lambda\gg v$. Within such a rather
general framework, we have employed a full set of $SU(2)\times
U(1)$ invariant dimension-six bosonic effective operators and
evaluated, at the one-loop level, the leading $O(1/\Lambda^2)$
correction to the well-known SM result. Such a calculation seems
to be missing in the earlier papers dealing with the $H\rightarrow
\gamma\gamma$ process. We have found out that the one-loop
contribution involving dimension-six bosonic operators can be very
important if there is an accidental suppression of the tree-level
effective interaction (i.e. when the combination
$f_{BB}-f_{BW}+f_{WW}$ is close to zero). More generally, it is
remarkable that the inclusion of dimension-six bosonic operators
can change dramatically the bosonic SM contribution; usually it is
expected that the effects of new physics beyond SM would enhance
the $H\rightarrow \gamma\gamma$ decay rate, but Fig.~\ref{gr1}
shows that the effects of the operator~(\ref{c6}) could also
reduce it significantly.

In our calculation we have ignored completely the effect of
fermions. Note that within SM the fermionic contribution to
$H\rightarrow\gamma\gamma$ is rather small for the light Higgs (at
least at one-loop level) in comparison with bosonic contribution
(this follows from straightforward evaluation of~(\ref{Fbos})
and~(\ref{Ffer})). Within effective Lagrangian approach, the
effects of higher-dimensional fermionic operators were studied
previously in the paper~\cite{hgg}. Needless to say, a complete
realistic calculation would have to take into account an educated
guess of the values of the coefficients $f_i$ (based on an
independent analysis of an appropriate set of other physical
processes); in this context see e.g.~\cite{zhang03} and references
therein. Further work in this direction is in progress.

\section*{Acknowledgments}
We are grateful to Ji\v{r}\'{\i} Novotn\'y for discussions. This
work was supported by Centre for Particle Physics, project No.
LN00A006 of the Ministry of Education of the Czech Republic.
\appendix
\section{Loop Functions}
In evaluating loop integrals we have used the so-called
Passarino-Veltman reduction~\cite{passarino79}, i.e. the reduction
of tensor one-loop integrals to the special scalar integrals which
can be further expressed by means of some standard analytical
functions. In the text we have employed the scalar integrals
$B_0$, $C_0$ and $D_0$ defined within dimensional regularization
scheme as
\begin{equation}
i\pi^2 B_0 (p_1^2,m_1^2,m_2^2) = \int\frac{d^d l}{(2\pi\mu)^{d-4}}
\frac{1}{[l^2-m_1^2][(l+p_1)^2-m_2^2]}\,,
\end{equation}
\begin{multline}
i\pi^2 C_0 (p_1^2,(p_1-p_2)^2,p_3^2;m_1^2,m_2^2,m_3^2)=\\
\int\frac{d^d l}{(2\pi\mu)^{d-4}}
\frac{1}{[l^2-m_1^2][(l+p_1)^2-m_2^2][(l+p_2)^2-m_3^2]}\,,
\end{multline}
\begin{multline}
i\pi^2 D_0 (p_1^2,(p_1-p_2)^2,(p_2-p_3)^2,p_3^2,p_2^2,(p_1-p_3)^2;m_1^2,m_2^2,m_3^2,m_4^2)=\\
\int\frac{d^d l}{(2\pi\mu)^{d-4}}
\frac{1}{[l^2-m_1^2][(l+p_1)^2-m_2^2][(l+p_2)^2-m_3^2][(l+p_3)^2-m_4^2]}\,.
\end{multline}
The $C_0$ and $D_0$ are UV finite while $B_0$ is UV divergent for
$d=4$. Defining
$$\frac{1}{\tilde\varepsilon} = \frac{2}{4-d} -
\gamma_E + \log 4\pi,
$$
we have
\begin{equation}
B_0(p^2,m^2,m^2) = \frac{1}{\tilde\varepsilon} -
\log\frac{m^2}{\mu^2} + 2 +{\scriptstyle\sqrt{1-4
\frac{m^2}{p^2}}} \log \frac{{\scriptstyle\sqrt{1-4
\frac{m^2}{p^2}}}-1}{{\scriptstyle\sqrt{1-4 \frac{m^2}{p^2}}}+1}.
\end{equation}
One of the three-point functions used in (\ref{fted}) is given by
\begin{multline}
C_0 (m_H^2,0,0;m_W^2) \equiv C_0 (m_H^2,0,0;m_W^2,m_W^2,m_W^2) =\\
-\frac{1}{m_H^2} \Bigl[ \text{Li}_2 \Bigl(\frac{2}{1-\sqrt{1- 4
m_W^2/{m_H^2}}}\Bigr) + \text{Li}_2 \Bigl(\frac{2}{1+\sqrt{1- 4
m_W^2/{m_H^2}}}\Bigr) \Bigr]\,,
\end{multline}
where $\text{Li}_2(x)$ is the standard dilogarithm defined through
the Spence's integral:
\begin{equation}
\text{Li}_2 (x) = -\int_0^x \frac{\log(1-t)}{t} dt.
\end{equation}
Further, let us denote
\begin{align*}
C_0 (m_H^2, m_H^2, 0;m_W^2) &\equiv C_0 (m_H^2,m_H^2,0; m_W^2,
m_W^2, m_W^2)\,,\\
D_0 (m_H^2; m_W^2) &\equiv D_0 (m_H^2, 0, 0, 0, 0, 0;
m_W^2,m_W^2,m_W^2,m_W^2)\,,\\
D_0 (m_H^2,m_H^2; m_W^2) &\equiv D_0 (m_H^2, m_H^2, 0, 0, 0, 0;
m_W^2,m_W^2,m_W^2,m_W^2)\,.
\end{align*}
A particular linear combination of these quantities appears in the
last two lines of eq.~(\ref{fted}). The resulting expression comes
out to be quite simple, namely
\begin{multline}
2m_W^2\bigl(D_0 (m_H^2;m_W^2) + 2 D_0(m_H^2,m_H^2;m_W^2)\bigr)
(m_H^2-6 m_W^2) - C_0(\ldots) (m_H^2+6 m_W^2)\\ = \frac{6(m_H^2 -
2 m_W^2)}{m_H\sqrt{4 m_W^2-m_H^2}} \text{arctan} \Bigl(
\frac{m_H}{\sqrt{4 m_W^2-m_H^2}}\Bigr)\,.
\end{multline}
Note finally that the proper analytic continuation of the above
functions is obtained by means of the prescription $m^2
\rightarrow m^2 - i \varepsilon$ wherever it is necessary.


\begin{thebibliography}{99}
\bibitem{barate03} R.~Barate {\it et al.} [ALEPH
Collaboration], %``Search for the standard model Higgs boson at LEP,''
Phys.\ Lett.\ B {\bf 565} (2003) 61 [arXiv:hep-ex/0306033].
\bibitem{sikivie80}
P.~Sikivie, L.~Susskind, M.~B.~Voloshin and V.~I.~Zakharov,
%``Isospin Breaking In Technicolor Models,''
Nucl.\ Phys.\ B {\bf 173} (1980) 189.
\bibitem{bsm}
L.~J.~Hall, {\it Beyond the standard model\/}, ICHEP 2000, Osaka,
Japan, 27 Jul - 2 Aug 2000.
\bibitem{wudka94}
J.~Wudka,
%``Electroweak effective Lagrangians,''
Int.\ J.\ Mod.\ Phys.\ A {\bf 9} (1994) 2301
[arXiv:hep-ph/9406205].
\bibitem{ns}
J.~Novotn\'y and M.~St\"ohr,
%``Anomalous couplings of vector bosons and the decay H $\to$ gamma gamma:  Dimensional regularization versus momentum cutoff,''
Czech.\ J.\ Phys.\  {\bf 49} (1999) 1471 [arXiv:hep-ph/9904401].
\bibitem{buchmuller85}
W.~Buchm\"uller and D.~Wyler,
%``Effective Lagrangian Analysis Of New Interactions And Flavor Conservation,''
Nucl.\ Phys.\ B {\bf 268} (1986) 621.
\bibitem{hagiwara93}
K.~Hagiwara, S.~Ishihara, R.~Szalapski and D.~Zeppenfeld,
%``Low-energy effects of new interactions in the electroweak boson sector,''
Phys.\ Rev.\ D {\bf 48} (1993) 2182.
\bibitem{cranmer04}
K.~Cranmer, B.~Mellado, W.~Quayle and S.~L.~Wu,
 %``Search for Higgs bosons decay H $\to$ gamma gamma using vector boson
%fusion,''
arXiv:hep-ph/0401088.
\bibitem{ellis75}
J.~R.~Ellis, M.~K.~Gaillard and D.~V.~Nanopoulos,
%``A Phenomenological Profile Of The Higgs Boson,''
Nucl.\ Phys.\ B {\bf 106} (1976) 292; M.~A.~Shifman,
A.~I.~Vainshtein, M.~B.~Voloshin and V.~I.~Zakharov,
%``Low-Energy Theorems For Higgs Boson Couplings To Photons,''
Sov.\ J.\ Nucl.\ Phys.\  {\bf 30} (1979) 711 [Yad.\ Fiz.\  {\bf
30} (1979) 1368].
\bibitem{konig91}
H.~K\"onig,
%``Higgs decay into two photons as a probe of anomalous gauge couplings,''
Phys.\ Rev.\ D {\bf 45} (1992) 1575.
\bibitem{hgg}
J.~M.~Hernandez, M.~A.~Perez and J.~J.~Toscano,
%``Decays H0 $\to$ Gamma Gamma, Gamma Z, And Z $\to$ Gamma H0 In The Effective Lagrangian Approach,''
Phys.\ Rev.\ D {\bf 51} (1995) 2044; M.~A.~Perez, J.~J.~Toscano
and J.~Wudka,
%``Two photon processes and effective Lagrangians with an extended scalar sector,''
Phys.\ Rev.\ D {\bf 52} (1995) 494 [arXiv:hep-ph/9506457].
\bibitem{banin98}
A.~T.~Banin, I.~F.~Ginzburg and I.~P.~Ivanov,
%``Anomalous interactions in Higgs boson production at photon colliders,''
Phys.\ Rev.\ D {\bf 59} (1999) 115001 [arXiv:hep-ph/9806515].
\bibitem{horejsi02}
J.~Ho\v{r}ej\v{s}\'\i: {\it Fundamentals of Electroweak Theory\/}
(Karolinum Press, Prague 2002).
\bibitem{bardin99}
D.~Y.~Bardin and G.~Passarino: {\it The Standard Model In The
Making: Precision Study Of The Electroweak  Interactions\/}
(Clarendon Press, Oxford 1999).
\bibitem{arason91}
H.~Arason, D.~J.~Castano, B.~Keszthelyi, S.~Mikaelian,
E.~J.~Piard, P.~Ramond and B.~D.~Wright,
%``Renormalization group study of the standard model and its extensions. 1. The Standard model,''
Phys.\ Rev.\ D {\bf 46} (1992) 3945.
\bibitem{zhang03}
B.~Zhang, Y.~P.~Kuang, H.~J.~He and C.~P.~Yuan,
%``Testing anomalous gauge couplings of the Higgs boson via weak-boson  scatterings at the LHC,''
Phys.\ Rev.\ D {\bf 67} (2003) 114024 [arXiv:hep-ph/0303048].
\bibitem{passarino79}
G.~Passarino and M.~J.~Veltman,
%``One Loop Corrections For E+ E- Annihilation Into Mu+ Mu- In The Weinberg Model,''
Nucl.\ Phys.\ B {\bf 160} (1979) 151.
\end{thebibliography}
\end{document}